\documentclass[pre,superscriptaddress,twocolumn]{revtex4-2}

\usepackage{siunitx}
\usepackage{color}
\usepackage[usenames,dvipsnames]{xcolor}
\usepackage{graphicx} 
\usepackage{tabularx} 

\definecolor{darkgreen}{rgb}{0.0, 0.6, 0.0}

\usepackage{soul}

\usepackage{marginnote}
\usepackage[perpage]{footmisc}

\usepackage{perpage}

\begin{document}

\title{Flow profiles near receding three--phase contact lines: Influence of surfactants}


\author{Benedikt B. Straub}
\affiliation{Max Planck Institute for Polymer Research, Ackermannweg 10, D--55128 Mainz, Germany}


\author{Henrik Schmidt}
\affiliation{Max Planck Institute for Polymer Research, Ackermannweg 10, D--55128 Mainz, Germany}

\author{Peyman Rostami}
\affiliation{Max Planck Institute for Polymer Research, Ackermannweg 10, D--55128 Mainz, Germany}
\affiliation{Leibniz-Institut f\"ur Polymerforschung, Hohe Straße 6, D--01069 Dresden, Germany}

\author{Franziska Henrich}
\affiliation{Max Planck Institute for Polymer Research, Ackermannweg 10, D--55128 Mainz, Germany}

\author{Massimiliano Rossi}
\affiliation{Department of Physics, Technical University of Denmark, DTU Physics Building 309, DK--2800 Kongens Lyngby, Denmark}
\affiliation{Institute of Fluid Mechanics and Aerodynamics, Bundeswehr University Munich, D--85577 Neubiberg, Germany}

\author{Christian J. K\"ahler}
\affiliation{Institute of Fluid Mechanics and Aerodynamics, Bundeswehr University Munich, D--85577 Neubiberg, Germany}

\author{Hans--J\"urgen Butt}
\affiliation{Max Planck Institute for Polymer Research, Ackermannweg 10, D--55128 Mainz, Germany}

\author{G\"unter K. Auernhammer}
\affiliation{Max Planck Institute for Polymer Research, Ackermannweg 10, D--55128 Mainz, Germany}
\affiliation{Leibniz-Institut f\"ur Polymerforschung, Hohe Straße 6, D--01069 Dresden, Germany}

\date{\today}

\begin{abstract}
The dynamics of wetting and dewetting is largely determined by the velocity field near the contact lines. 
For water drops it has been observed that adding surfactant decreases the dynamic receding contact angle even at a concentration much lower than the critical micelle concentration (CMC).
To better understand why surfactants have such a drastic effect on drop dynamics, we constructed a dedicated a setup on an inverted microscope, in which an aqueous drop is held stationary while the transparent substrate is moved horizontally.
Using astigmatism particle tracking velocimetry, we track the 3D displacement of the tracer particles in the flow. 
We study how surfactants alter the flow dynamics near the receding contact line of a moving drop for capillary numbers in the order of \num{e-6}. 
Even for surfactant concentrations $c$ far below the critical micelle concentration ($c \ll$ CMC) Marangoni stresses change the flow drastically. 
We discuss our results first in a 2D model that considers advective and diffusive surfactant transport and deduce estimates of the magnitude and scaling of the Marangoni stress from this.  
Modeling and experiment agree that a tiny gradient in surface tension of a few \si{\micro\newton\per\meter}  is enough to alter the flow profile significantly. 
The variation of the Marangoni stress with the distance from the contact line suggests that the 2D advection-diffusion model has to be extended to a full 3D model. 
The effect is ubiquitous, since surfactant is present in many technical and natural dewetting processes either deliberately or as contamination.
\end{abstract}


\maketitle

\section{Introduction}
The wetting and dewetting of a solid surface by a liquid are fundamental processes of many natural phenomena and technical applications. 
Examples include drops gliding down a window glass or being blown across a windscreen as well as drop and bubble motion in coating, painting, printing and flotation. 
Drops on a substrate are surrounded by a contact line, i. e., a line where substrate, liquid and gas  meet. 
The Young relation predicts a specific contact angle $\theta$ as a function of the surface tensions at the liquid-gas, liquid-substrate and substrate-gas interface \cite{Young:1805aa}.
For all practical purposes, the contact angle at non-moving contact lines is limited between the static advancing and receding contact angle due to imperfections of the substrate. 
This contact angle hysteresis give rise to a lateral adhesion force that has to be overcome to make a drop \cite{Gao:2017ab}. 
When the contact line is moving, the contact angle is velocity dependent. 
For simple liquids, the dynamics of wetting, i.e., the  velocity dependent advancing and receding contact angle, are well described by hydrodynamic modeling and molecular kinetic theory  \cite{moffatt64, huh71, voinov76, cox86, bonn_09, snoeijer_13}.
Essentially, the hydrodynamic energy dissipation is concentrated near the moving contact line, because of the apparently diverging shear rate at the moving contact line.  
The connection between the dissipation at the contact line and the macroscopic wetting dynamics near contact lines has been described using the Navier--Stokes equation. 
As Huh and Scriven pointed out, without further approximations, the dissipation near the receding contact line diverges \cite{huh71}. 
To circumvent this kind of unphysical singularities various ways of introducing cutoff lengths or slip boundary conditions on the solid substrate are discussed \cite{huh71, voinov76, cox86, bonn_09, snoeijer_13}.
In typical substrates the length scales where these effects become relevant are nanoscopic, i.e., inaccessible with optical measurement techniques \cite{Bonaccurso:2002aa}. 

As soon as the strong assumption of simple liquids (only one component, Newtonian viscosity, constant surface tension, etc.) is left, the modeling is much more complex, if existing \cite{lu_16}. 
In quasi-static situations surfactants change the surface tensions and consequently the contact angles of sitting drops. 
Under constant flow conditions, the presence of surfactants can change the boundary conditions on free liquid-gas interfaces. 
This is essentially due to the the fact that surface dilation or contraction due to hydrodynamic flow induces gradients in the surfactant surface concentration and consequently the surface tension. 
These gradients in surface tension, or Marangoni stresses, counteract the delating or contracting surface flows, i.e., modify the stress-free boundary condition at free surfaces. 
This is nicely illustrates in rising bubbles in surfactant solutions, which form stagnant caps even at very low concentrations  \cite{DAVIS1966681, Sadhal:1983aa, Dukhin:2015aa}, i.e., the boundary condition becomes effectively non-moving. 
These stagnant caps  change the rise velocity and flow field in and outside the bubble. 
A similar process is observed in the coalescence dynamics of drops \cite{amarouchene_01, manor_08, IASELLA2019136, Sauleda:2021us} with liquid pools. 
In evaporating drops, the natural convection due to thermal Marangoni effects \cite{Giorgiutti-Dauphine:2018vi} is modified for drops containing more than one component \cite{manikantan_squires_2020}, e.g., additional surfactants \cite{Anyfantakis:2015ab, marin_16}, particles \cite{Manukyan:2013aa}, or salt \cite{marin_19}. 

Landau-Levich films form above a critical speed, when plates are being pulled out of a liquid pool. 
The properties of these films and the hydrodynamic flow field in the pool depend on the surfactant concentration in the liquid \cite{Park:1991aa, KRECHETNIKOV:2006aa, mayer_12}.
When a continuous motion of the contact line is enforced, i.e., when at speeds below the critical speed for film formation, different scenarios have been observed. 
The presence of insoluble surfactants can induce an unsteady motion of the contact line \cite{Varanasi:2005aa} and structured deposits on the substrate \cite{Beppler:2008aa}. 

The situation can change qualitatively for soluble surfactants. 
Studies with a rotating drum setup show that even small amounts of soluble surfactants strongly reduce the dynamic receding contact angle \cite{fell_11,fell_13, henrich_16, truszkowska_17}.
By adding surfactant at a concentration between \num{5}--\num{30}\si{\percent} of the critical micelle concentration (CMC), the critical velocity for film formation reduces by one order of magnitude. 
This strong decrease in the receding contact angle was attributed to surface tension gradients (Marangoni effect) near the receding contact line, where continuously a new surface was assumed to be created. 

Measurements of the three--dimensional flow pattern of surfactant solutions near receding contact lines would shed light on the true mechanism behind this strong influence of small surfactant concentrations on the dynamic receding contact angle. 
For pure water, such measurements had been carried out by Kim et al. using a Tomographic PIV setup \cite{Kim_11, Kim_15}. 
Qian et al. measured the liquid velocities near the substrate in the vicinity of the contact lines using particle image velocimetry and flood illumination \cite{qian_15}. 
Since surfactants change the hydrodynamic boundary condition at the liquid-gas interface of liquids \cite{amarouchene_01, manor_08,chan_09}, surface velocimetry has been carried out. 
Studies using plates withdrawn from a liquid pool with surfactant solutions reveal a strong impact of the contact line velocity on the surface flow \cite{luokkala_01, beppler_08}. 
A recirculation flow on the surface was observed in dynamic wetting experiments of drops with insoluble surfactants \cite{leiske_11}. 
Yet, the internal dynamics can only be approximated indirectly with these methods. 
Other studies measured the internal flow during dewetting macroscopically \cite{henrich_16, mayer_12}. 
Due to resolution limits, these studies were not able to resolve the area near the contact line and liquid-gas interface. 

In this work we compare quantitatively measured flow profile near receding contact lines in surfactant solutions to modeling considerations of 2D models. 
In the following section, we describe a  advection-diffusion model for receding contact lines and explore some consequences thereof on the surface-tension gradients. 
The advection-diffusion model gives important predictions on the magnitude and scaling of the surface-tension gradients. 
We then report measurements of the flow profile and deduce the surface tension gradient close to receding contact line from these measurements. 
The comparison of the modeling part to the experiments leads us to the conclusion that a purely 2D approach is not compatible with the experimental data. 
Our results show that (i) minute  concentrations can influence the dynamics surfactant laden drops significantly and (ii) the dynamics of surfactant laden drops is a non-local process in which, most probably, the flow in the entire drop has to be considered. 

\section{Relevant mechanisms}
\label{sec:rel_mech}
\subsection{2D advection-diffusion model}
\subsubsection{General idea}
\label{sec:rel_mech:adv_diff}
\begin{figure}[htbp]
	\center
	\includegraphics[width=75.6mm]{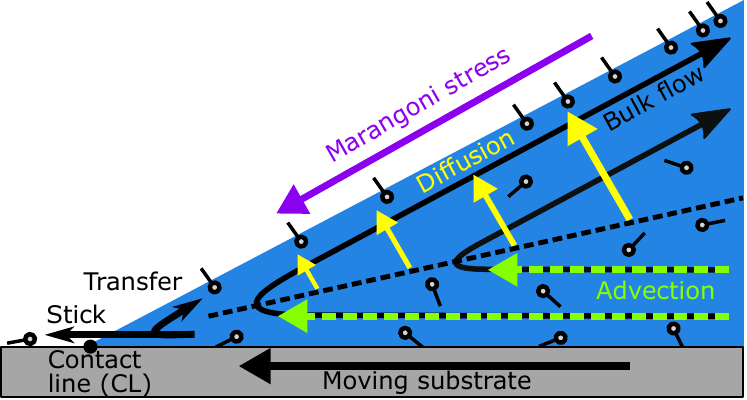}
	\caption{
	Simple 2D model to explain the Marangoni stress in the case of a receding contact line for aqueous surfactant solutions. A new liquid-gas interface is created at the contact line. Hence, the liquid surface is expanded. This expanded interface is initially less covered with surfactants (small dots with tail) than the surface in equilibrium. Below the dashed line, advection transports surfactant towards the contact line. Surfactants reach the expanded surface either by advection and diffusion away from the contact line or by direct transfer from the solid-liquid interface to the liquid-gas interface at the contact line. Part of the surfactants remain absorbed on the solid interface after the contact line passed. This leads to a concentration gradient, i.e., a Marangoni stress along the liquid-gas interface, which opposes the surface flow.
	} 
	\label{fig:1_Hypothesis}
\end{figure}

In a simple model that takes into account various previous studies \cite{fell_11,fell_13, henrich_16, truszkowska_17,amarouchene_01, manor_08,chan_09,luokkala_01, beppler_08,leiske_11,mayer_12, henrich_19, manikantan_squires_2020}, a two--dimensional model can be used as a working hypothesis (Figure \ref{fig:1_Hypothesis}): 
At the receding contact line new liquid-gas interface is created and thus the liquid surface is expanded. 
This expanded surface is less covered with surfactant than the liquid surface in equilibrium at the given surfactant concentration. 
Surfactants reach this interface in two ways: 
(1) Surfactants, that were adsorbed at the solid-liquid interface, can be directly transferred to the new liquid surface at the contact line. Otherwise, they remain on the solid interface. 
(2) Surfactants diffuse from the bulk solution towards the new surface, supported by advection due to the bulk flow. The equilibrium surfactant concentration in the vicinity of the solid is transported with the bulk flow towards the contact line.  
This equilibration process leads to a gradient in surface tension and thus a Marangoni stress along the interface. 
The Marangoni stress is directed towards the three--phase contact line and counteracts the flow along the liquid-gas  interface.
Higher surfactant concentrations should lead to higher Marangoni stress. 
The surface activity of the surfactant influences the Marangoni effect. 
Based on previous results by some of us \cite{fell_11, henrich_16, henrich_19}, we proposed that the Marangoni effect scales with the relative concentration, as measured in percent of  the CMC.

\subsubsection{Typical diffusion distance}
The wedge geometry has no intrinsic length scale. 
So the characteristic length scales of the system have to be deduced by the from the sample properties. 

\begin{figure}[htbp]
\begin{center}
\includegraphics[width=75.6mm]{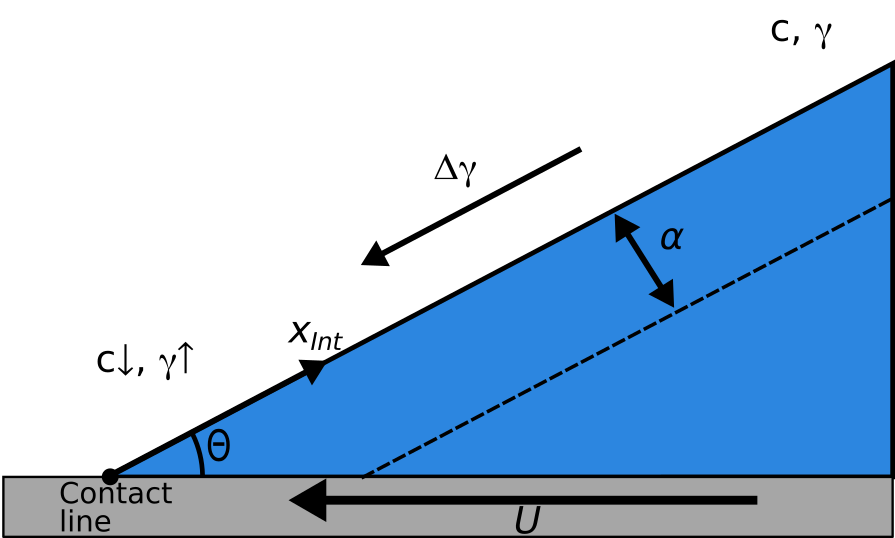}
\caption{Sketch of the area close to the moving three--phase contact line. }
\label{fig:geometry}
\end{center}
\end{figure}

To derive a characteristic diffusion distance, we compare the surface excess of the surfactants to their bulk concentration \cite{FERRI200061}. 
Using a linear dependence of the equilibrium surface excess $\Gamma$ on the bulk concentration $c$ \cite{fell_11, henrich_19}
\begin{equation}
\Gamma = \alpha c.
\label{eq:Linearisierung}
\end{equation}
The parameter $\alpha$ has the dimension of a length. 
We note that Equation \ref{eq:Linearisierung} is only valid for small concentrations.
In equilibrium, the length $\alpha$ is the thickness of a liquid layer underneath the liquid-gas interface that contains as many surfactant molecules as the liquid-gas interface,  Figure \ref{fig:geometry}. 
The length $\alpha$ can be calculated by using the Gibbs adsorption isotherm 
\begin{equation}
\Gamma = - \frac{c}{RT} \frac{d\gamma}{dc}.
\end{equation}
Here,  $R$ is the ideal gas constant and $T$ is the absolute temperature.
Figure \ref{fig:geometry} defines the orientation of the length $\alpha$.
By integrating the Gibbs adsorption isotherm between $c = 0$ and CMC of the surfactant, one gets the length $\alpha$
\begin{equation}
\alpha = \frac{\gamma_0 - \gamma}{2RTc}.
\end{equation}
Here, $\gamma_0$ is the surface tension of pure water (\SI{72.4}{\milli\newton\per\meter}) and $\gamma$ is the equilibrium surface tension of the surfactant solution at CMC. 
Since the liquid layer with the thickness $\alpha$ contains as many surfactant molecules as the surface, $\alpha$ is a good measure of a typical diffusion length in the simple 2D advection-diffusion model above.
For surfactants that give a similar surface tension $\gamma$ at very different absolute concentrations, i.e. for surfactants with strongly different CMC, the length scale $\alpha$ can vary significantly. 

\subsubsection{Typical advected distance}
In the 2D advection-diffusion system described above, the transport across $\alpha$ is purely diffusive. 
This allows us to estimate first a diffusion time $t_D$. 
\begin{equation}
t_D = \frac{\alpha^2}{2D},
\end{equation}
with the diffusion constant of the surfactant $D$. 

\begin{table*}[htbp]
\centering
\caption{Critical micelle concentration (CMC), surface tension at CMC $\gamma$, diffusion coefficient $D$ at 25 \si[mode=text]{\celsius} \cite{henrich_16}, the characteristic diffusion ($\alpha$) and advection distances ($x_{adv}$) of two typical non-ionic surfactants.}
\begin{tabularx}{1\textwidth}{>{\raggedright\arraybackslash}p{3.5cm} >{\raggedright\arraybackslash}X >{\raggedright\arraybackslash}X >{\raggedright\arraybackslash}X >{\raggedright\arraybackslash}X >{\raggedright\arraybackslash}X >{\raggedright\arraybackslash}X }
\hline
Surfactant  & Formula & CMC [\si[mode=text]{\milli M}] &  $\gamma$ [\si[mode=text]{\milli\newton\per\meter}] & D [\si[mode=text]{\meter\squared\per\second}] &  $\alpha$ [\si[mode=text]{\micro \meter}] & $x_{adv}$ [\si[mode=text]{\micro \meter}]\\
\hline
\rule{0pt}{3ex}Octyl triglycole & $\mathrm{C_8E_3}$ & 7.5 & 27.3 & \num[mode=text]{4.6e-10} & 2.5 & 1.2\\
\rule{0pt}{3ex}Dodecyl pentaglycole & $\mathrm{C_{12}E_5}$ & 0.07 & 30.7 & \num[mode=text]{2.9e-10} & 244 & 6900\\
\hline
\end{tabularx}
\label{tab:surfactant_data}
\end{table*}
 
 To estimate the advected distance during this diffusion time, we use the velocity of the liquid-gas interface $U_{int}$ in the surfactant-free case as derived by Moffatt \cite{moffatt64} 
\begin{equation}
U_{int} = U \frac{\sin \theta - \theta \cos \theta}{\theta - \sin \theta \cos \theta}.
\end{equation}
Consequently, the advected distance is given by 
\begin{equation}
x_{adv} = t_D U_{int}.
\end{equation}
As the estimates given in Table \ref{tab:surfactant_data} indicate, different surfactants can have huge differences in these characteristic length scales. 
We will check these dependences experimentally in the later sections of this work. 

\subsection{Gradients along the liquid-gas interface in a 2D flow}
\label{sec:Model_grad}
A first hint that this simple 2D advection-diffusion model might have to be extended came from a set of experiments in the rotating drum setup \cite{fell_13}. 
The absolute magnitude of the changes in the dynamic contact angle depends strongly on the possible transport mechanism of the surfactant between the receding  and the advancing contact line. 
For example, blocking the transport along the liquid-gas interface between the advancing and the receding contact line changes significantly the magnitude of the change in the dynamic contact angles \cite{fell_13}. 

Further limitation of this simple model becomes apparent, when considering the stress balance at the free surface close to the receding contact line. 
The surface tension gradient $\nabla_{\parallel} \gamma$ due to the adsorption of the surfactants on the freshly formed liquid-gas interface has to be balanced by the viscous stress of the hydrodynamic flow $\tau = \mu \nabla_{\perp} \vec{v}_p$, with $v_p$ the flow component parallel to the liquid-gas interface $\mu$ the viscosity of the liquid and $\nabla_{\parallel}$ and $\nabla_{\perp}$ the gradient parallel and perpendicular to the liquid-gas interface 
\begin{equation}
\tau = \mu \nabla_{\perp} \vec{v}_p = \nabla_{\parallel} \gamma.
\label{eqn:shear_stress}
\end{equation}

To get an estimate of the upper bound of the gradients in surface tension, we assume that a gradient is strong eneough to stop the surface flow at the liquid-gas interface completely.
Such situation is solved analytically by Taylor in what is typically called a scraping flow \cite{Taylor:1960uv, Batchelor:2000wd}. 
The tangental stress along the liquid-gas interface needed to stop the flow is given by \cite{wikipediea_taylor_scraping}
\begin{equation}
\mu \nabla_{\perp} \vec{v}_p = \frac{2 \mu U}{x_{int}}\frac{\sin \theta - \theta \cos \theta}{\theta^2 - \sin^2 \theta}, 
\end{equation}
where $\theta$ is the contact angle in radiants and  $x_{int}$ the distance to the contact line (Figure \ref{fig:geometry}). 
Integrating this from infinity to the a finite distance from the contact line gives an upper bound for the expected differences in the surface tension needed to completely stop the surface flow away from the contact line 
\begin{equation}
\Delta \gamma = -2 \mu U\ln{x_{int}} \frac{\sin \theta - \theta \cos \theta}{\theta^2 - \sin^2 \theta}.
\label{eq:upper_bound_d_gamma}
\end{equation}

Importantly, in this 2D case the change in surface tension has only a weak logarithmic dependence on the distance from the contact line. 
To get an estimate of the expected changes, we enter typical values for the experimental parameters into Equation \ref{eq:upper_bound_d_gamma} ($U = \SI{200}{\micro \meter \per \second}$, $\mu = \SI{1}{\milli\pascal\second}$, $\theta = \pi/4$ , and $x_{int} = \SI{50}{\micro \meter}$), see also the experimental part below. 
With these numbers, we get a surface tension difference in the order of $\approx \SI{5}{\micro\newton\per\meter}$. 
This extremely small value is essentially due to the low viscosity of water. 

\subsection{Dimensionless numbers}

Besides the estimates of characteristic length, time and surface tension scales as given above, we use dimensionless numbers to choose a good set of experimental settings. 
In the present case, several dimensionless numbers are connected to the problem. 
The Laplace number $La$ measures the ratio of the capillary forces to the inertial forces in a fluid flow. 
The Laplace number is also the ratio of the Reynolds $Re$ and the capillary number $Ca$. 
Both numbers give dimensionless velocities. 
The Reynolds number relates the fluid inertia to the viscous dissipation. 
It is often used to characterize the crossover from laminar to turbulent flow. 
The capillary number give the ratio of viscous forces to surface tension forces: 
\begin{equation}
La = \frac {\gamma \rho L}{\mu^2} = \frac {Re}{Ca} 
\label{Eq:Laplace_num}
\end{equation}
\begin{equation}
Re = \frac {u \rho L}{\mu} 
\label{Eq:Reynolds_num}
\end{equation}
\begin{equation}
Ca = \frac{\mu u}{\gamma}
\label{Eq:Capillary_num}
\end{equation}
In these definitions $\rho$ is the mass density, $L$ a characteristic length scale, and $u$ a characteristic velocity. 

Additionally, one further dimensionless number could play a role: the Damk\"ohler number \cite{manikantan_squires_2020}. 
This is given by  the ratio between the diffusion time to the interface and the adsorption time at the interface.  
However, previous studies have shown that the overall effect on the dynamic receding contact angle is  independent on the charge and molar mass of the surfactant \cite{henrich_16, truszkowska_17}. 
These studies concluded that the rate limiting step is not the adsorption to the liquid-gas interface but the transport to the interface. 
We follow this result and assume in this work that the adsorption process of the surfactant from the subsurface to the surface is fast compared transport process to the (sub)surface. 

The typical number given in Table \ref{tab:dimensionless_num} show important characteristics of the flow of surfactant solutions close to receding contact lines. 
The flow is laminar ($Re \ll 1$) and dominated by capillary effects ($La \gg 1$ and  $Ca \ll 1$). 
These relations stay valid, if the length $\alpha$ ist taken as a characteristic length. 
\begin{table}[htbp]
\centering
\caption{Estimates of typical dimensionless numbers: Laplace number $La$, Reynolds number $Re$, and Capillary number $Ca$, with a characteristic length $L$ of \SI{50}{\micro \meter}, the viscosity of water $\mu = \SI{1}{\milli\pascal\second}$, the density of water $\rho =  \SI{1000}{\kilo\gram\per\meter^3}$, and a characteristic velocity of \SI{200}{\micro\meter \per \second}. }
\begin{tabularx}{1\columnwidth}{>{\raggedright\arraybackslash}p{3.5cm} >{\raggedright\arraybackslash}X >{\raggedright\arraybackslash}X >{\raggedright\arraybackslash}X }
\hline
Surface tension & $La$ & $Re$ &  $Ca$\\
\hline
\rule{0pt}{3ex}\SI{72}{\milli\newton\per\meter} & $3600$ & $0.01$ & $2.8 \times 10^{-6}$\\
\rule{0pt}{3ex}\SI{40}{\milli\newton\per\meter} & 2000 & 0.01 & $5.0 \times 10^{-6}$ \\
\hline
\end{tabularx}
\label{tab:dimensionless_num}
\end{table}

\subsection{Questions to the experiments}

The above considerations give a couple of non-trivial predictions on the characteristics of the flow field in surfactant solutions close to receding contact lines that can be checked experimentally: 
(i) Close to the receding contact line, a region with a slowed down or stagnated surface flow should be observed. 
(ii) The size of this region should depend strongly on the type of the surfactant, especially, its values of the characteristic length $\alpha$.
 (iii) The absolute value of the change in surface tension needed to produce these effects is very small, in the range of several \si{\micro\newton \per \meter}. 
(iv) The surface tension should vary slowly with a logarithmic dependence on the distance to the contact line. 

To test these modeling observations experimentally, we varied two important parameters: (i) the characteristic length $\alpha$  by choosing two surfactants with different CMC (or the absolute surfactant concentration at comparable surface tensions) and (ii) by changing the concentration of these surfactant to test for a possible dependence on the Laplace number (or the concentration relative to CMC).

\section{Material and Methods}

 \begin{figure*}[htbp]
	\center
	\includegraphics[width=116mm]{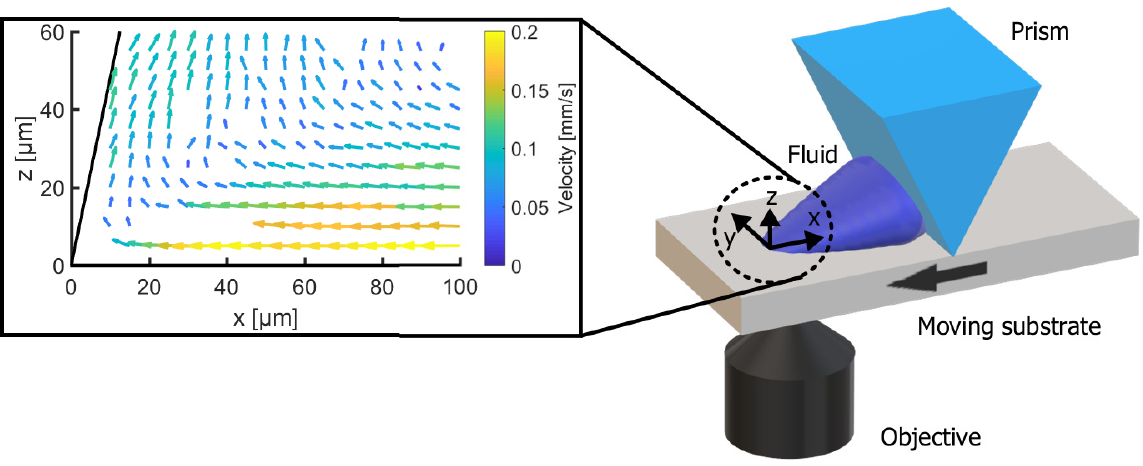}
	\caption{
	Sketch of the experimental setup. The shown flow field for pure water was measured close to the receding three--phase contact line and is shown in the co--moving frame of the contact line. The measured particle velocities within a volume around each grid point were averaged. The liquid close to the substrate has the highest velocity, directed parallel to the moving substrate and decreasing with increasing distance from the substrate. When getting closer to the contact line the flow is directed upwards. Contact line velocity: \SI{200}{\micro\meter\per\second}; receding contact angle: \ang{78}\ang{\pm 3}.
	} 
	\label{fig:2_Setup}
\end{figure*}

\subsection{Sample preparation}
We study solutions of nonionic surfactants, to avoid charge effects. The surfactants dodecyl pentaglycole ($\mathrm{C_{12}E_5}$)
and octyl triglycole ($\mathrm{C_8E_3}$),
purchased from Sigma--Aldrich, were used without further purification. The properties of the surfactants are listed in Table \ref{tab:surfactant_data}.

To track the flow motion we use fluorescent polystyrene particles (diameter \SI{2}{\micro\meter} or \SI{4}{\micro\meter}) dispersed in the aqueous surfactant solutions. The particles were either synthesized in house (Rhodamine B: Ex./Em. \SI{550}{\nano\meter}/\SI{600}{\nano\meter}) or bought from microParticles GmbH (PS--FluoRed: Ex./Em. \SI{530}{\nano\meter}/\SI{607}{\nano\meter}). To avoid sedimentation of the particles a 1:1 mixture of water ($\mathrm{H_{2}O}$, Arium\textregistered 611 or Arium\textregistered pro VF/UF \& DI/UV ultra--pure water systems, Sartorius, resistivity of \SI{18.2}{\mega\ohm}) and deuterated water ($\mathrm{D_{2}O}$ purchased from Sigma--Aldrich, purity \SI{99}{\percent}) was used.
Previous experiments have already demonstrated that these particles can faithfully follow the internal and interfacial flow of droplets \cite{marin_16, rossi2019interfacial}. Precision cover glasses (No. 1.5H, $24 \times 60$ \si{\milli\meter\squared}) were mechanically cleaned with acetone, isopropanol, ethanol using cleanroom wipes. Afterwards, the cover glasses were hydrophobized with trichloro(1H,1H,2H,2H--perfluorooctyl)silane (Sigma--Aldrich) via the gas phase. The cover glasses were placed side by side with two \SI{10}{\micro\liter} drops of the silane in a desiccator for \SI{2}{\minute}. A magnetic ventilator ensured a homogeneous distribution of the silane in the gas phase.

\subsection{Experimental setup}
To image the moving contact line for a long time and with high resolution, the drop was pinned to a prism while the substrate was moved smoothly back and forth (Figure \ref{fig:2_Setup} (a)) by a piezoelectric motor (LPS--45, Physik Instrumente, Germany). 
We placed this setup above the objective lens of an inverted fluorescence microscope. 
Thus, the moving contact line was stationary as seen by the microscope. 
The distance between the prism and the measurement area was at least a factor of \num{30} larger than the measurement area. 
Therefore, we assumed that the measured flow fields are independent of any influence of the prism. 
In this work, we considered receding contact lines with a constant velocity of \SI{200}{\micro\meter\per\second}, i.e., the drop was pulled away from the prism (Figure \ref{fig:2_Setup}). 
A stable steady--flow condition was achieved \SI{5}{\second} after starting the substrate motion and could be held for more than \SI{60}{\second}.
\begin{figure*}[tbp]
 \centering
  \includegraphics[width=139.3mm]{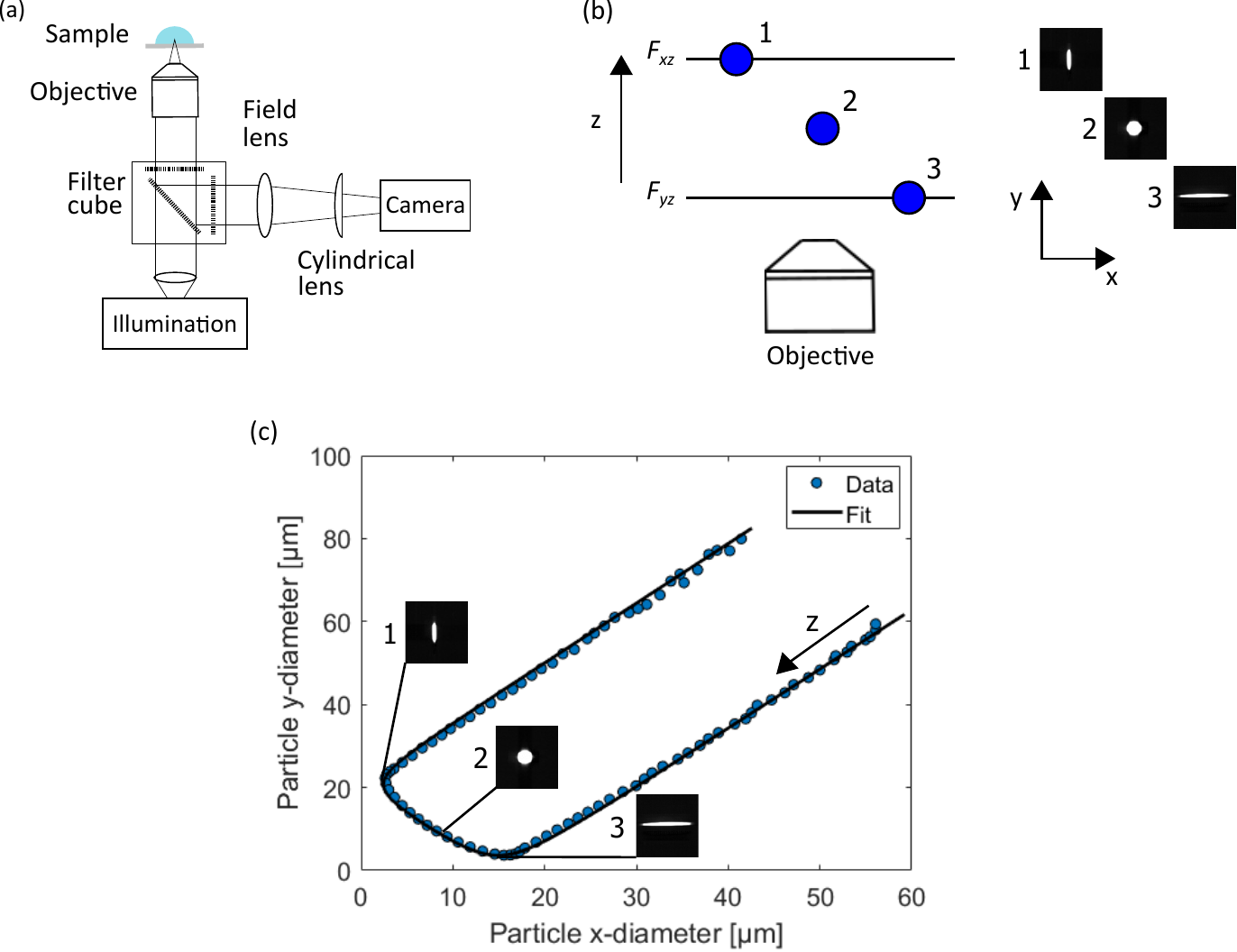} 
    \caption{
    (a) Sketch of an APTV setup with a cylindrical lens in front of the camera chip. (b) Working principle of APTV.
	The image of a fluorescent labelled particle changes depending on its distance to the objective lens, i.e., the $z$ position.
	(c) The scatter plot of $x$-- and $y$--extensions of particle images at different $z$ positions collapses on a single curve that is dependent on the axial variable $z$.
	This curve is obtained by a calibration procedure on reference particle images. The calibration curves are used to calculate the $z$ position of measured particle images.
    }
 \label{fig:3_APTV_Working_Principle}
\end{figure*}

\subsection{Astigmatism particle tracking velocimetry}
Three--dimensional particle trajectories were measured by astigmatism particle tracking velocimetry (APTV). 
APTV is a single--camera method that uses a deliberate astigmatic aberration of the optical system, obtained by introducing a cylindrical lens in front of the camera sensor \cite{kao94, cierpka11, rossi14}. 
Figure \ref{fig:3_APTV_Working_Principle} shows a sketch of the used APTV setup. 
The working principle of APTV is shown in Figure \ref{fig:3_APTV_Working_Principle} (b). 
The optical configuration of an APTV system has not one, but two principal focal planes ($F_{xz}$ and $F_{yz}$).
The reason for that is the cylindrical lens in front of the camera chip, which introduces a controlled astigmatic aberration.
The particle image deformation depends on the relative distance of the particle to the focal axes.
In the example in Figure \ref{fig:3_APTV_Working_Principle} (b), particle 1 is on the $F_{xz}$ plane and therefore it is focused in $x$--direction ($x$--diameter small) and defocused in $y$--direction ($y$--diameter large).
On the other hand, particle 3 is on the $F_{yz}$ plane and therefore it is focused in $y$--direction and defocused in $x$--direction.
If the $x$-- and $y$--diameters of particle images at different axial positions are plotted together, they collapse on a single curve as a function of the axial particle position (Figure \ref{fig:3_APTV_Working_Principle} (c)).
This curve can be obtained by a calibration procedure performed on reference particle images at known axial positions.
The axial position of a measured particle image is then determined by comparing its $x$ and $y$ diameters with the calibration curve.
In our analysis we follow the  APTV procedure as described in \cite{cierpka11, rossi14}.

For the APTV measurements, we used a \SI{40}{X} microscope objective in combination with a \SI{150}{\milli\meter} cylindrical lens. 
Images were acquired with a rate of \SI{50}{\hertz}. 
The maximum measurement volume of the setups was around ${400 \times 450 \times 60}$\si{\,\micro\cubic\meter} with uncertainty in the particle position determination of less than \SI{\pm 0.25}{\micro\meter} in the lateral direction and \SI{\pm 1}{\micro\meter} in the axial direction. 
Detailed information about the setups and data processing can be found in the \mbox{Supplemental Material \cite{SI}}.

The velocity fields were obtained from an average of 3 experiments for each fluid solution.
In each experiment, an average of 168 tracer particles were tracked and the respective positions and velocities measured with APTV.
The position and inclination of the substrate was determined experimentally from the trajectories of particles stuck on the substrate or trapped at the contact line.
All data are shown in a coordinate system in which the plane at $z = 0$ corresponds to the substrate.
The position of the contact line was also identified in each image and used to correct the stream--wise coordinate $x$, so that the final $x$--coordinate measures the distance of particles from the contact line (with $x = 0$ corresponding to the contact line).
Finally, all data points were projected on the $xz$--plane.
The measured particle velocities within a volume around each grid point were averaged with an an average number of 13 particle velocities for each grid point.

\section{Measured characteristic of the flow field}
\subsection{Velocity and deviation fields}
\begin{figure*}[htbp]
 \centering
  \includegraphics[width=172mm]{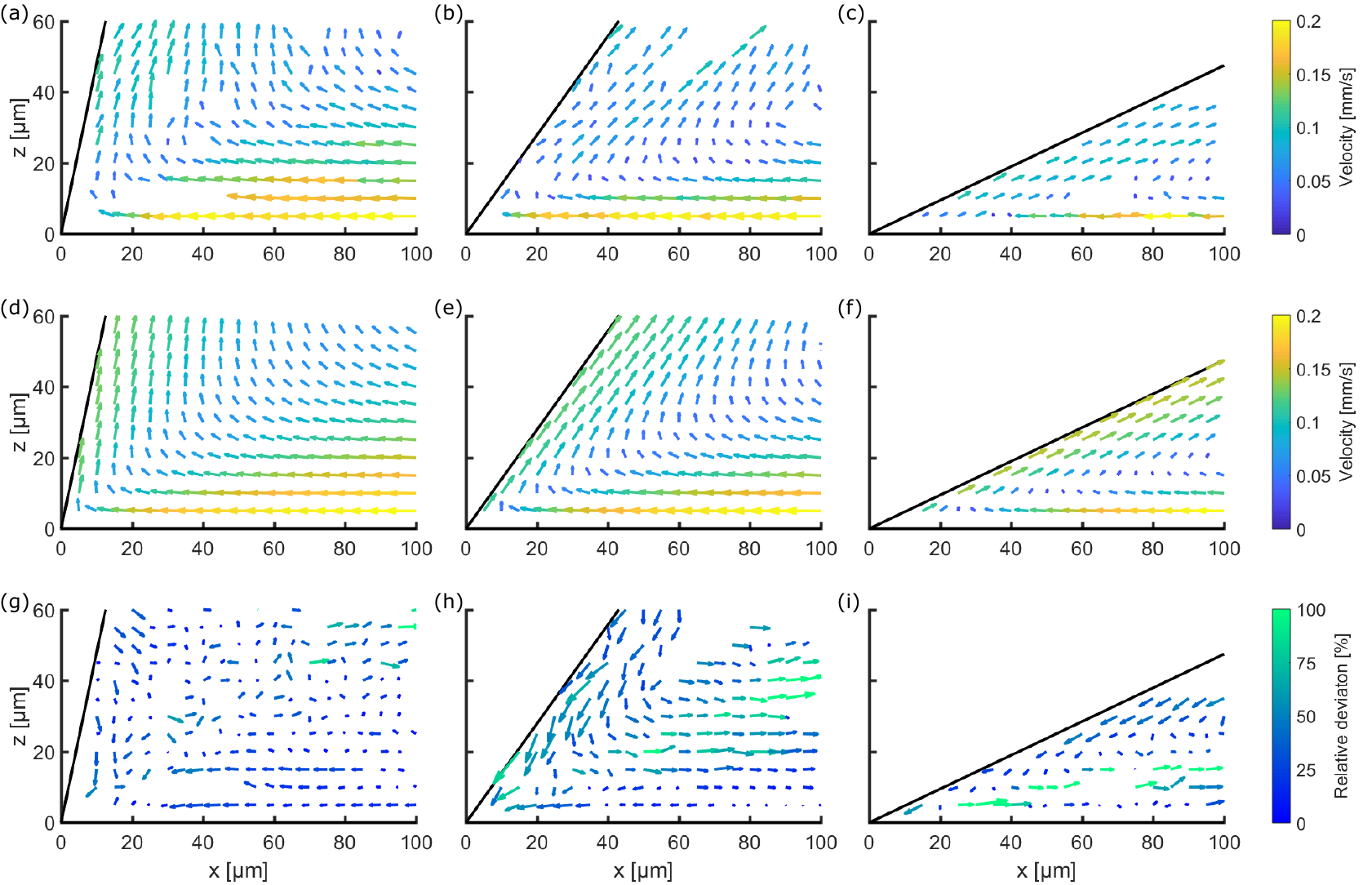} 
    \caption{
    Direct comparison between the measured velocity fields (a--c), the analytical solution of Moffatt \cite{moffatt64} (d--f) and the corresponding deviation fields (g--i). 
	Each column represents one measurement solution and the contact line velocity is always \SI{200}{\micro\meter\per\second}.
	The first column is the measured flow field of pure water (a), (d) is the corresponding analytical solution and (g) is the calculated deviation field between (a) and (d).
	The second column shows the measured flow field for a \SI{30}{\percent CMC} $\mathrm{C_{12}E_5}$ solution (b) and the calculated deviation field (h).
	The third column shows the results for a \SI{30}{\percent CMC} $\mathrm{C_8E_3}$ solution.
	The comparison between the measured flow fields and the analytical solution shows that in the case of surfactant solutions, the flow at the interface between liquid and air is reduced.
This is shown clearly by the deviation fields (h, i).
   }
 \label{fig:3_Deviation_fields}
\end{figure*}
To compare experimental flow profiles (e.g. Figure \ref{fig:3_Deviation_fields} (a--c)) to hydrodynamic calculations, we assume a simplified geometry near the contact line: 
Rather than treating the full three--dimensional dewetting problem, we take a two-dimensional wedge geometry, because we do not observe significant transverse flow within the resolution of our measurements. 
For this geometry, hydrodynamic flow is known for simple liquids, like water \cite{moffatt64, huh71,voinov76, cox86}. 
These theories provide an analytical solution for the region very near the three--phase contact line (inner solution) and the bulk flow (outer solution). 
Due to the size of the tracer particles, we could not resolve experimentally the flow in the inner region. 
Therefore, we compare our measurements only to the outer theoretical solution, which is the same for all these theories.  
We use the theory of Moffatt \cite{moffatt64} for stress--free liquid-gas interface and constant contact line velocity. 
Figure \ref{fig:3_Deviation_fields} (g) depicts the deviation field between measurement (a) and theory (d) for pure water. 
This deviation field does not show a flow field but the normalized velocity difference between measurement and theory as calculated by:
\begin{equation}
u_{d} = u_{m} - u_{t},\quad w_{d} = w_{m} - w_{t},
\label{eqn:deviation_field}
\end{equation}
\begin{equation}
S_d = \frac{|\vec{v}_d|}{|\vec{v}_t|} = \frac{\sqrt{u_d^2+w_d^2}}{\sqrt{u_t^2+w_t^2}} .
\label{eqn:deviation_field_2}
\end{equation}
Here, $u$ and $w$ are the mean local velocity components in $x$ and $z$ direction at a given ($x$, $z$) position. 
The indices indicate deviation $d$, measurement $m$ and theory $t$. 
The relative strength of the deviation $S_d$ is represented by the color code and the length of the arrows, the angular deviation by the orientation of the arrows. 
The deviation field of water (Figure \ref{fig:3_Deviation_fields} (g)) has a randomly distributed angular deviation and its magnitude is \SI{<30}{\percent}. 
Due to this randomness and the small absolute deviation, measurement and theory agree very well. 
The random noise shows that no bias errors are present. 
At the liquid-gas interface, little deviation occurs. 
As expected, the liquid-gas interface is stress--free.

In the presence of surfactants (\SI{30}{\percent CMC}), the receding contact angle at a velocity of \SI{200}{\micro\meter\per\second} decreased from \ang{78} (pure water) to \ang{54} ($\mathrm{C_{12}E_5}$, \SI{0.38}{ppm}) and \ang{25} ($\mathrm{C_8E_3}$, \SI{40}{ppm}). 
Because of this change in contact angle, it is impossible to directly compare the measured flow fields of surfactant solutions to those measured in pure water. 
No hydrodynamic theory is available to describe the flow of surfactant solutions near moving contact lines. 
As shown above, the water measurements match the hydrodynamic theory for water. 
So, we can compare the surfactant measurements with the hydrodynamic theory for pure water \cite{moffatt64} at the respective contact angle.

For both surfactants, the region near the liquid-gas interface has a reduced velocity; the deviation field points towards the contact line (Figure \ref{fig:3_Deviation_fields} (h), (i)). 
The surface flow is slower than in pure water. 
The surface velocity is reduced in the entire observation volume near the free surface. 
As will be shown in the next section, this reduction in surface velocity results from a Marangoni stress caused by gradients in surfactant concentration, which opposes the surface flow.

If we consider the shown fluid volume of Figure \ref{fig:3_Deviation_fields} (h) as a control volume, we can apply the conservation of mass. 
Fluid enters the volume at \mbox{$x\,=\,$\SI{100}{\micro\meter}} from the right and leaves the volume at \mbox{$z\,=\,$\SI{60}{\micro\meter}} parallel to the liquid-gas interface. 
Since the outflow of the fluid near the free surface is reduced due to the Marangoni stress, the inflow of the fluid into the volume is also reduced to fulfil mass conservation; the deviation field points to the right at \mbox{$x\,=\,$\SI{100}{\micro\meter}}. 
The same general qualitative behavior is observed for surfactants with very difference $\alpha$, showing this does not depend on the characteristic length scale $\alpha$ but is a generic feature.

\subsection{Deduced changes in surface tension}
To quantify the gradient in surface tension and surface excess, i.e., the surface concentration of surfactant molecules, along the free surface, we use the already mentioned Equation (\ref{eqn:shear_stress}). 
We first calculate the shear stress from the measured velocity profiles. 
Since any tangental stress occurring at a liquid-gas interface originates from a surface tension gradient, we calculate the surface tension gradient along the free surface $\nabla_{\parallel} \gamma$ from the shear stress $\tau = \mu \nabla_{\perp} \vec{v}_p$ in the flow profile just under the free surface \cite{landau_87}, Equation (\ref{eqn:shear_stress}):
$ \tau = \mu \nabla_{\perp} \vec{v}_p = \nabla_{\parallel} \gamma $. 
The velocity $\vec{v}_p$ is the velocity component parallel to the free surface. 

\begin{figure}[htbp]
	\center
	\includegraphics[width=75.6mm]{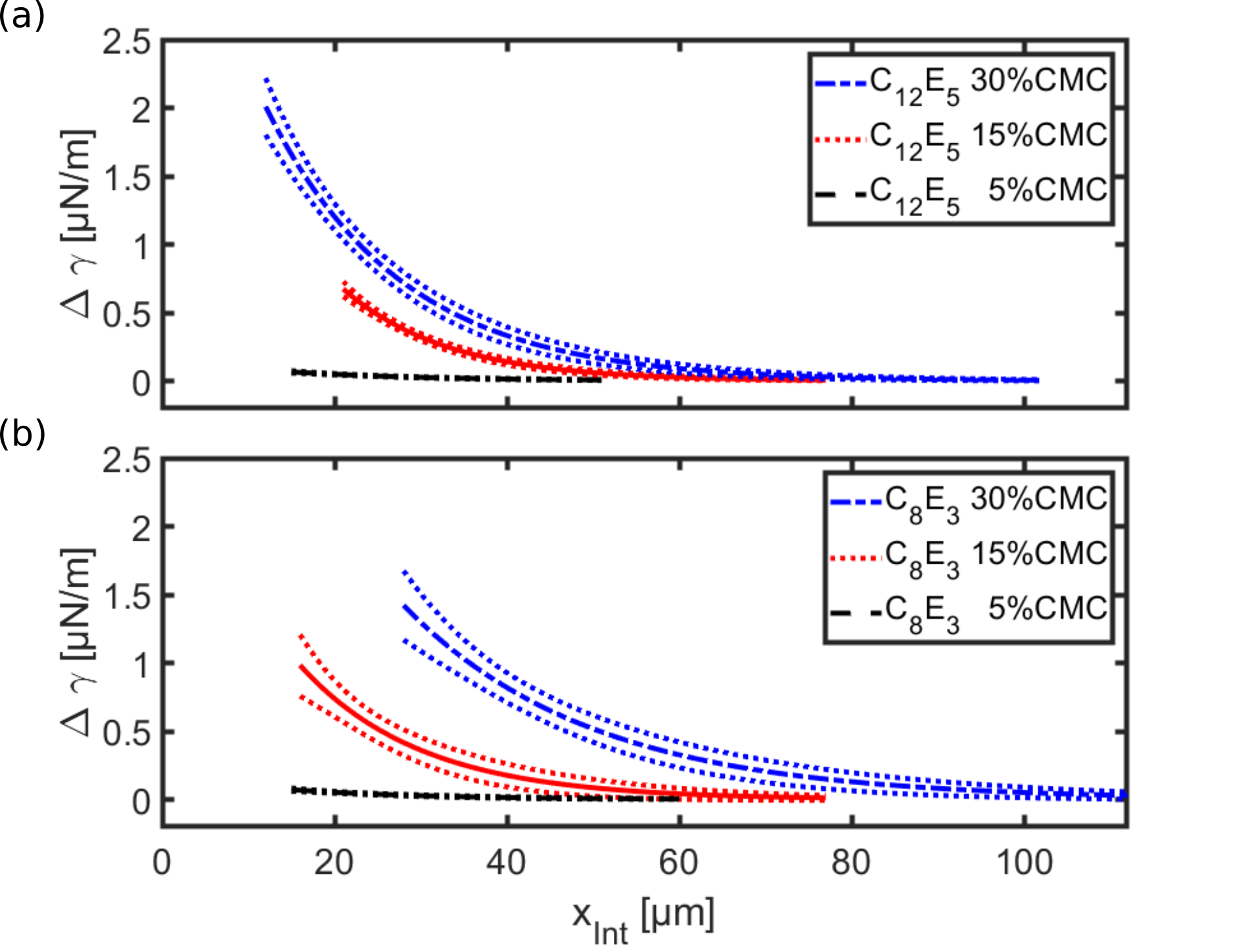}
	\caption{
	(a) and (b) show the calculated change in surface tension $\Delta \gamma$ for different concentrations of $\mathrm{C_{12}E_5}$ and $\mathrm{C_8E_3}$ along the free surface $x_{Int}$. See Fig.~\ref{fig:geometry} for the axis direction.
	The data was fitted with an exponential function and the dotted lines are the corresponding prediction bounds. 
	} 
	\label{fig:4_Delta_Gamma}
\end{figure}

To calculate the velocity derivative, we fit the velocity in the direction normal to the free surface by a cubic polynomial. 
We differentiate the polynomial function to calculate the shear stress $\tau$. 
We use the viscosity of water $\mu = \SI{1}{\milli\pascal\second}$ to calculate the stress for the different solutions. 
Due to the scattering of the experimental data the uncertainty of the calculated stress is around \SI{20}{\percent}. 
Since the statistical error becomes too large closer than \SI{20}{\micro\meter} to the contact line, we exclude this region from further analysis. 
The surface tension gradient $\nabla_{\parallel} \gamma$ is integrated from right to left to obtain the change in surface tension $\Delta \gamma$ (Figure \ref{fig:4_Delta_Gamma}).

For all measured concentrations, a clear surface tension gradient was measurable.
Far away from the contact line (\SI{>70}{\micro\meter}) the surface--tension gradient vanishes.
At this distance the gradient in surface concentration is too small to produce measurable effects, i.e., we consider the surface tension to be close to constant. 
When approaching the contact line, the surface tension deviates from its equilibrium value. 
For both surfactants the magnitude in surface tension difference $\Delta \gamma$ increases with increasing surfactant concentration, Figure \ref{fig:4_Delta_Gamma}. 
Although the change in surface tension for the \SI{5}{\percent CMC} solutions is  small, it is significantly larger than our resolution limit that is given by the apparent surface tension gradient for water, see Supplemental Material \cite{SI}.  
We quantify this range of the increased surface tension by a decay length $L_D$ over which $\Delta \gamma$ reduces to a fraction of $1/e$ of its initial value. 
The decay lengths of the surface tension gradient differ by less than a factor of two between the used surfactants and concentrations, Table \ref{tab:surfactant_transport_process}.
Additionally, there is no clear tendency of  how the decay length depends on the characteristic length $\alpha$, i.e., on the type of the surfactant, or on the Laplace number $La$, i.e., on the balance of the surface forces to the viscous forces in the liquid. 
So neither molecular nor local dynamic properties seem to play a dominating role in defining the hydrodynamic flow close to the receding contact line in these surfactant solutions. 

Although the surface tension near the contact line is only increased by \mbox{\num{1}--\num{2}$\,$\si{\micro\newton\per\meter}}, the gradient in surface tension, i.e. the Marangoni stress, is around \mbox{\num{70}--\num{130}$\,$\si{\micro\newton\per\meter\squared}}. 
This increase in surface tension corresponds to a decreasing surface concentration, more precisely surface excess, of surfactant near the contact line. 

\begin{table*}[htbp]
	\centering
	\caption{
	Solution parameters: relative concentration in \si{\percent CMC}, absolute concentration in the bulk $c$ in parts per million, equilibrium surface tension $\gamma$, equilibrium surface excess $\Gamma$, surface area per molecule $A_{mol}$, length $\alpha$ and decay length $L_D$ for different surfactant concentrations. 
	}
	\label{tab:surfactant_transport_process}
	\begin{tabularx}{1\textwidth}{ >{\raggedright\arraybackslash}X  >{\raggedright\arraybackslash}X  >{\raggedright\arraybackslash}X  >{\raggedright\arraybackslash}X  >{\raggedright\arraybackslash}X  >{\raggedright\arraybackslash}X  >{\raggedright\arraybackslash}X >{\raggedright\arraybackslash}X}
		\hline
		\rule{0pt}{3ex}Name & [\si{\percent CMC}] & $c$ [\si{ppm}] & $\gamma$ [\si{\milli\newton\per\meter}] & $\Gamma$ [\si{\mol\per\meter\squared}] & $A_{mol}$ [\si{\nano\meter\squared}] & $\alpha$ [\si{\micro\meter}] & $L_D$ [\si{\micro\meter}] \\
		\hline
		\rule{0pt}{3ex}$\mathrm{C_{12}E_5}$ & \num{5} & \num{0.06} & \num{58.2} & \num{0.9e-6} & \num{1.9} & \num{244} & \num[separate-uncertainty = true,mode=text]{17.5+-3.9}\\
		\rule{0pt}{3ex}$\mathrm{C_{12}E_5}$ & \num{15} & \num{0.2} & \num{47.5} & \num{2.6e-6} & \num{0.7} & \num{244} & \num[separate-uncertainty = true,mode=text]{12.4+-1.6}\\
		\rule{0pt}{3ex}$\mathrm{C_{12}E_5}$ & \num{30} & \num{0.4} & \num{40.2} & \num{5.1e-6} & \num{0.3} & \num{244} & \num[separate-uncertainty = true,mode=text]{15.8+-2.1}\\
		\hline
		\rule{0pt}{3ex}$\mathrm{C_{8}E_3}$ & \num{5} & \num{6.8} & \num{53.2} & \num{0.9e-6} & \num{1.8} & \num{2.5} & \num[separate-uncertainty = true,mode=text]{17.4+-4.8}\\
		\rule{0pt}{3ex}$\mathrm{C_{8}E_3}$ & \num{15} & \num{20.3} & \num{43.4} & \num{2.8e-6} & \num{0.6} & \num{2.5} & \num[separate-uncertainty = true,mode=text]{15.9+-5.5}\\
		\rule{0pt}{3ex}$\mathrm{C_{8}E_3}$ & \num{30} & \num{40.7} & \num{35.4} & \num{5.6e-6} & \num{0.3} & \num{2.5} & \num[separate-uncertainty = true, mode=text]{23.3+-6.0}\\
		\hline
	\end{tabularx}
\end{table*}

\section{Comparing experimental data to the modeling ideas}
\subsection{Gradients in surface tension}
The measured flow fields and absolute values match to some extend the expected values. 
As discussed in Section \ref{sec:Model_grad}, the magnitude of the change in surface tension is in the order of \si{\micro\newton\per \meter}. 
This is actually expected due to the close relation between the magnitude of this gradient and the viscosity of the liquid. 
For the same reason, the surface excess near the contact line is only around \SI{0.01}{\percent} lower than the equilibrium surface excess. 
These small changes of surface excess strongly influence the flow profile and the apparent receding contact angle (Figure \ref{fig:3_Deviation_fields}) at very small absolute concentrations of the surfactants at \SI{30}{\percent CMC} of only \SI{0.38}{ppm} and \SI{40}{ppm} for $\mathrm{C_{12}E_5}$ and $\mathrm{C_8E_3}$, respectively. 
This observation indicates that even minor contamination or small amounts of additives may strongly influence the flow near contact lines. 

Our experimental results give clear data on the flow field close to the contact line and consequently on the gradients in surface tension and surface excess. 
We can, however, draw no conclusions on the amount of surfactant transferred from the solid-liquid interface to the liquid-gas interface directly at the moving contact line, because we lack data in the last $\approx 20$ \si{\micro \meter}. 
Our data shows that the surface is also expanded in our observation region. 
The surface velocity is increasing with increasing distance from the contact line over a distance that is roughly given by the decay length $L_D$, Figure \ref{fig:3_Deviation_fields}.
Since we lack data very close to the receding contact line, we are unable to quantify the true surface generation directly at the receding contact line. 

There are some substantial differences of the experimental data to the modeling expectations. 
(i) The decay in surface tension is much faster than the logarithmic decay predicted by the simple 2D model. 
(ii) For all concentrations and both surfactants, the decay lengths $L_D$ of the surface tension gradients are almost the same (Table \ref{tab:surfactant_transport_process}). 
(ii) This measured decay length matches none of the expected advected lengths $x_{adv}$. 
For  $\mathrm{C_8E_3}$ the decay length is much larger than the advected length $x_{adv} \ll L_D$, for  $\mathrm{C_{12}E_5}$ the inequality is inverted $x_{adv} \gg L_D$.
For a precise comparison one would have to take the actual surface velocities in calculating the advected length. 
However, this does not resolve the discrepancies because changing the surface velocity by maximum a factor of three over a few tens of micrometers does not change the advected distance to the amount needed to identify it with the decay length $L_D$. 

These discrepancies clearly show the limitations of the 2D modeling considerations in Section \ref{sec:rel_mech}.
Especially, neither the properties summarized in the characteristic length $\alpha$ (molecular properties of the surfactant)  nor the Laplace number $La$ (characteristic of the dynamics close to the receding contact line) seem to have major effect. 

From this we conclude that one of the essential approximation of the simple model has to be dropped. 
The advection-diffusion approach is a simple consequence of mass conservation and has a strong basis. 
In contrast the 2D approximations has a much weaker basis and has already be shown to have some limitations \cite{fell_13}. 
In the following we explorer possibilities beyond this 2D approximation. 

\subsection{Beyond the 2D approximation}

\begin{figure}[h]
	\center
	\includegraphics[width=84.3mm]{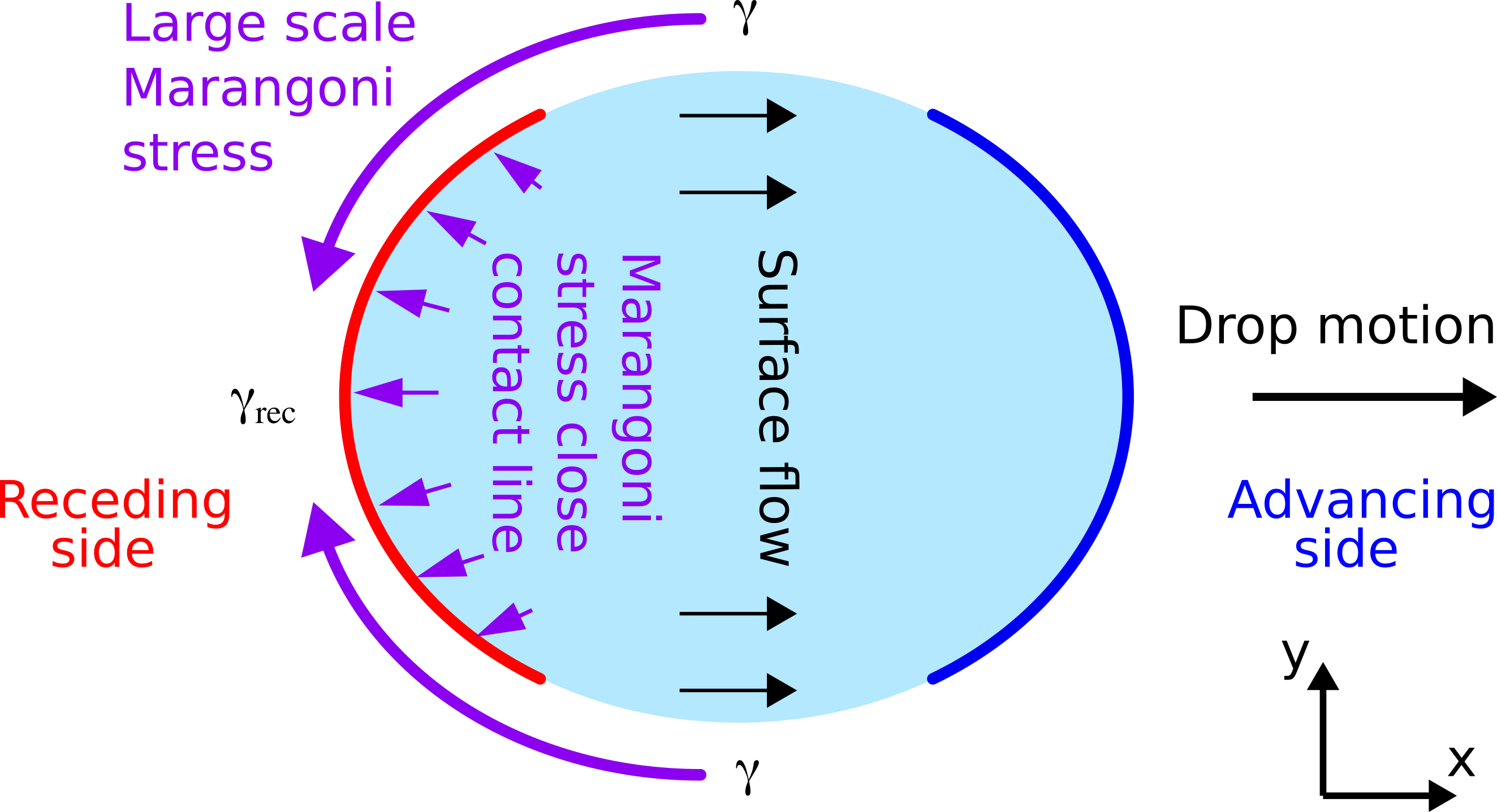}
	\caption{
Sketch of the three--dimensional Marangoni stress around the drop's surface (purple arrows). 
The black arrows in the drop indicate the surface flow. 
At the sides of the drop, between the areas of the advancing and receding contact lines,  no liquid-gas interface is generated and thus the surface tension is closer to equilibrium and higher than close to the receding contact line. 
We measure that the surface tension close the receding side is reduced (Figure \ref{fig:4_Delta_Gamma}). 
Hence, a surface tension gradient along the drop's surface exists that can drive  surface transport of surfactant from the sides of the drop to the receding end of the drop.
Note the different length scales. The width of the drop is several \si{\milli\meter} and the region with lower surface tension close to the receding contact line has an extension in flow direction of several tens of \si{\micro \meter}. 
	} 
	\label{fig:three_dimensional_flow}
\end{figure}

As shown in a previous study \cite{fell_13}, the surface transport of the surfactant has an important influence on the dynamic receding contact angles. 
When taking the entire drop into account, also a fully 3D contribution becomes apparent. 
Close to the receding side of the drop, the generation rate for new liquid-gas interface is highest and the liquid surface is measurably expanded, Figure \ref{fig:3_Deviation_fields}. 
At the side of the drop, between the advancing and receding end of the drop, there is a region in which the surface flow in the drop is parallel to the contact line, as sketched in Figure \ref{fig:three_dimensional_flow}. 
The further the liquid-gas interface is away from the receding contact line, or the ''older'' this surface is, the closer the surface concentration is to the equilibrium surface concentration. 
Thus, at the sides of the drop, where no fresh liquid-gas interface is generated, the surface concentration of surfactant is higher than close to the receding contact line. 
Correspondingly, there is a surface tension gradient, a large scale Marangoni stress, parallel to the contact line that can drive  surface transport of surfactant from the drop sides to the receding end of the drop.
This Marangoni stress can modify the over all flow profile in the drop, leading to an additional transport of surfactant to the region  close to the receding contact line. 
In this 3D model, the spatial distribution of surface excess of surfactant is also determined by the entire flow profile in the drop and not by the local processes close to the receding contact line, i.e., thus a fully 3D flow allows to resolve the discrepancies described in the previous section. 

\section{Summary and conclusion}

We have analyzed the flow field in the vicinity of the receding contact line of a water drop with and without surfactant. 
The measured flow profiles support the idea that receding contact angles decrease even at low surfactant concentration because of a Marangoni effect. 
Near the receding contact line the concentration of surfactant is reduced leading to a Marangoni stress, i.e., gradient in surface tension. 

Two system parameters are important in a 2D model considering advective and diffusive transport of surfactant to the free surface. 
(i) The ratio between surface excess and volume concentration of the surfactant give a characteristic diffusion length $\alpha$. 
(ii) The dimensionless Laplace number $La$ gives the ratio between capillary forces and inertial forces. 
The characteristic numbers $\alpha$ and $La$ are varied by using different surfactants and different concentrations. 
The 2D advection-diffusion model gives some strong predictions on the characteristics of the flow field, namely the order of magnitude of the surface tension gradient and its dependence on the distance to the contact line. 
The flow velocity and viscosity of the aqueous drops studied only leads to a very small gradient in surface tension and surface excess in the observed volume. 
The experimentally measured flow fields allow to check these predictions. 
The magnitude of the measured surface tension gradient fits nicely in the predicted range of a few \si{\micro\newton\per\meter}. 
In contrast, the spatial dependence does not match the 2D advection-diffusion model, but decays much faster and does not depend on $\alpha$ and $La$. 
To resolve this, we propose a 3D flow field that is beyond the 2D approximation initially assumed. 
This conclusion is supported by the fact that the range of the surface tension gradients seems to be independent of the type of the surfactant and its concentration, i.e., of $\alpha$ and $La$. 

The dynamics of surfactant laden drops shows a strong coupling of the internal hydrodynamic flow and the surfactant dynamics at the liquid-gas interface. 
This coupling can limit the applicability of 2D approximations significantly. 
Our results imply that even small pollutions or additives may play a major role in dynamic dewetting processes.

\begin{acknowledgments}
Funded by the Deutsche Forschungsgemeinschaft (DFG, German Research Foundation) --- Project--ID 265191195 --- SFB 1194, A02 \mbox{(P. R.),} A06 \mbox{(B. B. S.,} \mbox{G. K. A.)} and C03 \mbox{(H.J. B.)}.
\mbox{H. S.} acknowledges financial support from the Deutsche Forschungsgemeinschaft (DFG, German Research Foundation) through Project No. AU321/3 within the Priority Program 1681.
\mbox{C. J. K.} and \mbox{M. R.} acknowledge financial support through Project No. KA1808/22--1.
The research leading to these results has received funding from the European Research Council under the European Union's Seventh Framework Programme (FP7/2007-2013)/ERC grant agreement n$^\circ$883632 (H.J. B., B. B. S.).
\end{acknowledgments}

\bibliography{Manuskript_Flow_profiles_near_moving_three_phase_contact_lines_Influence_of_surfactants}

\end{document}



\title{Flow profiles near \black{receding} three--phase contact lines: Influence of surfactants\\ Supplemental Material}


\author{Benedikt B. Straub}
\affiliation{Max Planck Institute for Polymer Research, Ackermannweg 10, D--55128 Mainz, Germany}


\author{Henrik Schmidt}
\affiliation{Max Planck Institute for Polymer Research, Ackermannweg 10, D--55128 Mainz, Germany}

\author{Peyman Rostami}
\affiliation{Max Planck Institute for Polymer Research, Ackermannweg 10, D--55128 Mainz, Germany}
\affiliation{Leibniz-Institut für Polymerforschung, Hohe Straße 6, D-01069 Dresden, Germany}

\author{Franziska Henrich}
\affiliation{Max Planck Institute for Polymer Research, Ackermannweg 10, D--55128 Mainz, Germany}

\author{Massimiliano Rossi}
\affiliation{Department of Physics, Technical University of Denmark, DTU Physics Building 309, DK--2800 Kongens Lyngby, Denmark}
\affiliation{Institute of Fluid Mechanics and Aerodynamics, Bundeswehr University Munich, D--85577 Neubiberg, Germany}

\author{Christian J. Kähler}
\affiliation{Institute of Fluid Mechanics and Aerodynamics, Bundeswehr University Munich, D--85577 Neubiberg, Germany}

\author{Hans--Jürgen Butt}
\affiliation{Max Planck Institute for Polymer Research, Ackermannweg 10, D--55128 Mainz, Germany}

\author{Günter K. Auernhammer}
\affiliation{Max Planck Institute for Polymer Research, Ackermannweg 10, D--55128 Mainz, Germany}
\affiliation{Leibniz-Institut für Polymerforschung, Hohe Straße 6, D-01069 Dresden, Germany}

\date{\today}


\maketitle

\vfill
\tableofcontents 

\renewcommand{\theequation}{S\arabic{equation}}
\renewcommand{\thefigure}{S\arabic{figure}}
\renewcommand{\thetable}{S\arabic{table}}
\renewcommand{\thepage}{\roman{page}}
\renewcommand{\bibnumfmt}[1]{[S#1]}
\renewcommand{\citenumfont}[1]{S#1}


\newpage
\section{Astigmatism Particle Tracking Velocimetry}
We performed the experiments in two facilities, setup 1 at the Max Planck Institute for Polymer Research in Mainz, and setup 2 at the Institute of Fluid Mechanics and Aerodynamics at the Bundeswehr University Munich. We obtained similar results in both experiments. 
\black{The final plots for pure water, $\mathrm{C_{12}E_5}$ and \SI{5}{\percent CMC} $\mathrm{C_8E_3}$ were obtained from setup 1 and the remanining $\mathrm{C_8E_3}$ plots from setup 2.}

\begin{figure*}[h]
	\center
	\includegraphics[width=170.62mm]{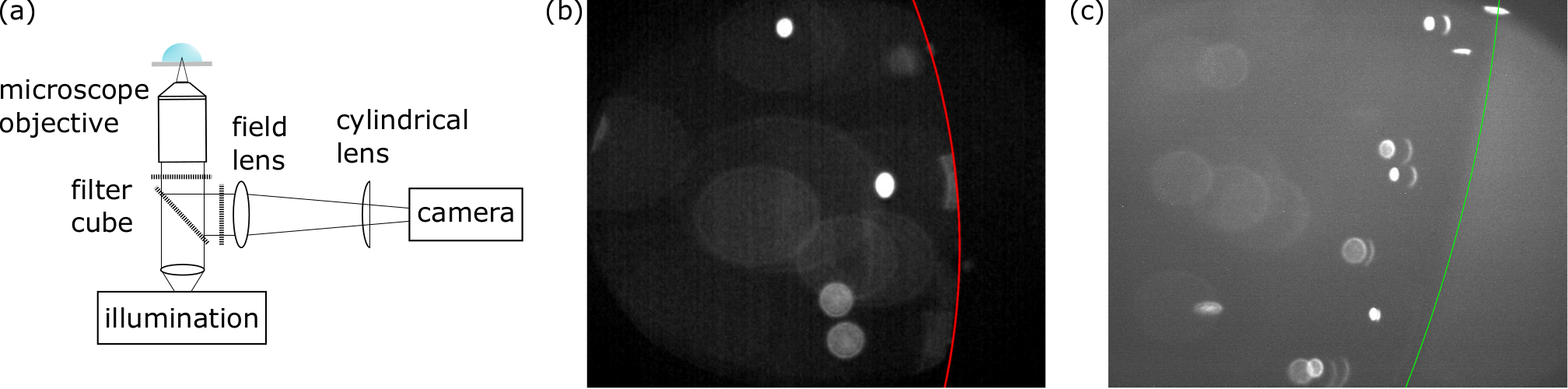}
	\caption{
	(a) Sketch of the experimental setups.
	(b, c) Examples of astigmatic particle images obtained with setup 1 and 2, respectively.
	The contact line of the drop is highlighted in red or green.
	} 
	\label{fig:2_Setup_1_schematic}
\end{figure*}

The  setup 1 consisted of a Leica DMI 6000B inverted microscope in combination with a Photron Fastcam 1.1.
Images were recorded with a recording speed of \SI{50}{\hertz} which provided good spatial and temporal resolution.
The optical elements were a \SI{150}{\micro\meter} achromatic cylindrical lens (Thorlabs) directly in front of the camera sensor  and a 40X microscope objective (LUCPLFLN, Olympus, Japan).
Illumination was provided by a mercury lamp in combination with an appropriate filter cube to adjust the used wavelength.
As tracer particles, red--fluorescent polystyrene spheres with a diameter of 2 \si{\micro\meter} were used (microParticles GmbH, PS--FluoRed).  
Small polystyrene beads are slightly denser than water ($\rho_{\mathrm{PS}} = 1050$ kg/\si{\cubic\meter}).
Therefore, to avoid sedimentation o he particles a 1:1 mixture of water and deuterated water was.
\black{Furthermore, the sedimentation time of the particles (even in pure water) is much larger than the time scales of the phenomena investigated, therefore it can safely be assumed that they follow faithfully the fluid flow \cite{rossi2019interfacial}.}
Figure \ref{fig:2_Setup_1_schematic} (a) shows a schematic sketch of the setup. 
The maximum measurement volume of this setup configuration is around $400 \times 450 \times 60\,\si{\micro\cubic\meter}$. 

The detection of the particle position and shape is made in two steps.
The first step is a pre--detection of the particles using a binarized image.
In the second step the pre--detection information is used as the start parameter for an iterative two--dimensional Gaussian fitting procedure to the full intensity profile of the particle image. 
See \cite{cierpka11} for the used Gaussian fit function.
The fit parameters of the Gaussian function are the extensions in both direction and middle point of the tracer particles.
To minimize errors the fitting is done iteratively with varying segmentation sizes.

To calibrate the system several particles are placed on top of a glass cover slip.
A drop of the measurement fluid is placed above the particles without disturbing them.
The focus of the microscope objective is moved through the probe with known step sizes.
The result is a calibration curve like shown in FIG. 3 (c) of the main article.
The fit function for the principal axes with the dependent axial variable $z$ can be found in \cite{cierpka11}.

The particle coordinates are combined to particle trajectories by using the “track” algorithm of Crocker and Grier \cite{crocker_96}.
Due to the cylindrical lens the image is also globally deformed.
Therefore, the lateral particle coordinates have to be transformed back to an undeformed image.
We took pictures of a standard calibration grid (PS20, Pyser Optics, Kent, UK) before and after the introduction of the cylindrical lens into the beam path.
We used the "bUnwarpJ" \cite{sorzano_05, carreras_06} Plugin of ImageJ \cite{schindelin_12, schneider_12, rueden_17} to generate a transformation matrix.
We used this matrix to transform the lateral coordinates of the particles back into the undeformed system.

The error estimation was carried out using test data.
These test data were measured the same way as the calibration curves.
The axial position of the particles is therefore known.
The calibration is used on this data and the deviation between the known axial position and the calculated axial position with the calibration is used as the axial error.
The axial standard deviation is \SI{\pm 0.98}{\micro\meter} and the estimated  lateral uncertainty is \SI{\pm 0.25}{\micro\meter}.\\

The  setup 2 was similar to setup 1 but used different components.
Specifically, it consisted of an inverted microscope Axio Observer Z1 (Carl Zeiss AG) in combination with an Imager sCMOS camera (LaVision GmbH), with images also recorded at \SI{50}{\hertz}.
The optical elements were a \SI{150}{\micro\meter} cylindrical lens place also in front of the camera sensor and a \SI{40}{X} microscope objective (LD Plan–Neofluar, Carl Zeiss AG).
The illumination was provided by a high-power green light--emitting diode (LED).
As tracer particles, red--fluorescent (Rhodamine B) polystyrene spheres with a diameter of \SI{4}{\micro\meter} were used (synthesized at the Max Planck Institute for Polymer Research). 

To obtain the calibration images for setup 2, a similar procedure as the one described for setup 1 was used.
Instead, a different approach was used to obtain the axial coordinate.
Rather than comparing the $x$ and $y$ extensions, the shape of the measured particle images was directly compared with the shape of the reference particle images by means of the normalized cross--correlation function.
More details on this calibration approach are given in \cite{barnkob2015general}.
Also in this case, a standard calibration grid (Thorlabs) was used to map the image distortion and get the coordinates in real units. 

The measurement volume obtained with this configuration was about $380 \times 430 \times 70\, \si{\cubic\micro\meter}$, with an estimated uncertainty of \SI{\pm 0.7}{\micro\meter} in the axial direction, and \SI{\pm 0.1}{\micro\meter} in the lateral direction.

%
%
%
%
\section{Velocity and deviation fields}

\begin{figure*}[h]
	\center
	\includegraphics{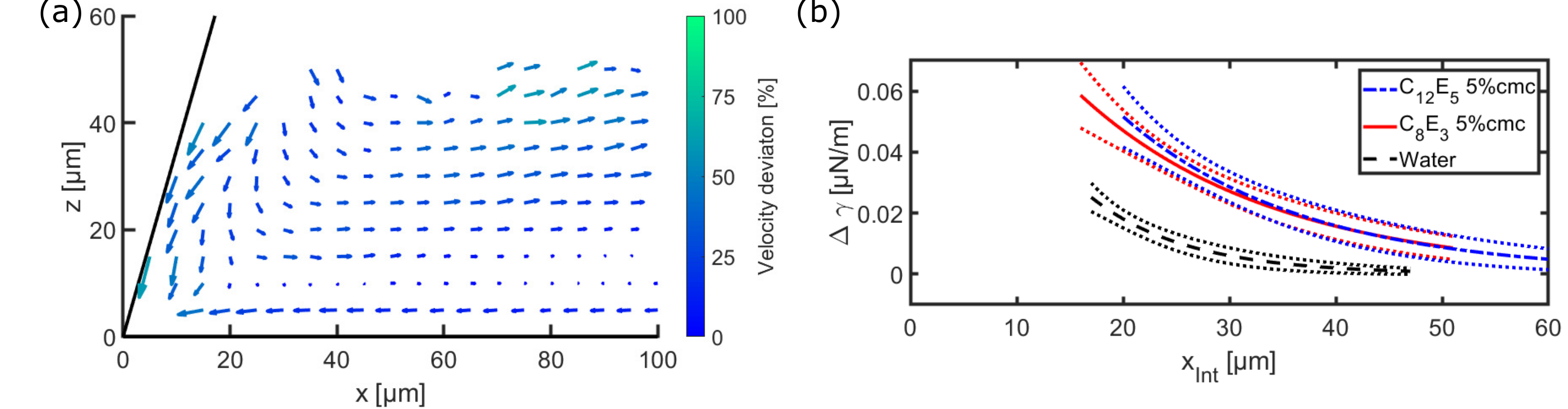}
	\caption{
	(a) Deviation field of \SI{5}{\percent CMC} $\mathrm{C_{12}E_5}$ solution. (b) Change in surface tension of water, \SI{5}{\percent CMC} $\mathrm{C_{12}E_5}$ and \SI{5}{\percent CMC} $\mathrm{C_8E_3}$.
	} 
	\label{fig:4_5cmc_water}
\end{figure*}

\black{
The deviation field of a \SI{5}{\percent CMC} $\mathrm{C_{12}E_5}$ also shows the discussed deviation close the the free surface and in the bulk flow (Fig. \ref{fig:4_5cmc_water} (a)).
Although the change in surface tension is very small, it is still a factor 3 higher than the noise level we measured for pure water.
We attribute the small systematic variation of the calculated surface tension difference for water to effect of the finite size of our tracer particles. 
Note that this systematic error has a different dependence on the distance from the contact line and falls off faster. 
}

\newpage

\bibliography{SI_Flow_profiles_near_moving_three_phase_contact_lines_Influence_of_surfactants}